\documentclass[sigconf]{acmart}

\usepackage{booktabs} 
\usepackage{amsmath}
\usepackage{caption}
\usepackage{subfigure}

\begin{document}
	\title{LogCanvas: Visualizing Search History Using Knowledge Graphs}

	\author{Luyan Xu}
	\affiliation{%
		\institution{DEKE lab\\ Renmin University of China}
		\streetaddress{Zhongguancun Street 59}
		\city{Beijing}
		\state{China}
		\postcode{100872}
	}
	\email{xuluyan@ruc.edu.cn}
	
	\author{Zeon Trevor Fernando}
	\affiliation{%
		\institution{L3S Research Center\\
			Leibniz Universit\"at Hannover}
		\streetaddress{Appelstr. 9a}
		\city{Hannover}
		\state{Germany}
		\postcode{30167}
	}
	\email{fernando@L3S.de}
	
	\author{Xuan Zhou}
	\affiliation{%
		\institution{School of Data Science \& Engineering\\ East China Normal University}
		\streetaddress{Zhongshan N. Rd 3663}
		\city{Shanghai}
		\state{China}
		\postcode{200062}}
	\email{zhou.xuan@outlook.com}
	
	\author{Wolfgang Nejdl}
	\affiliation{%
		\institution{L3S Research Center\\			Leibniz Universit\"at Hannover}
		\streetaddress{Appelstr. 9a}
		\city{Hannover}
		\state{Germany}
		\postcode{30167}}
	\email{nejdl@L3S.de}
	
	\renewcommand{\shortauthors}{Luyan Xu, Z.T.Fernando et al.}

	\begin{abstract}
		In this demo paper, we introduce LogCanvas, a platform for user search history visualization. 
		Different from the existing visualization tools, LogCanvas focuses on helping users re-construct the semantic relationship among their search activities. 
		LogCanvas segments a user's search history into different sessions and generates a knowledge graph to represent the information exploration process in each session. 
		A knowledge graph is composed of the most important concepts or entities discovered by each search query as well as their relationships. It thus captures the semantic relationship among the queries. 
		LogCanvas offers a session timeline viewer and a snippets viewer to enable users to re-find their previous search results efficiently. LogCanvas also provides a collaborative perspective to support a group of users in sharing search results and experience.
	\end{abstract}
	
	%
	%


	
	\copyrightyear{2018} 
	\acmYear{2018} 
	\setcopyright{rightsretained} 
	\acmConference[SIGIR '18]{The 41st International ACM SIGIR Conference on Research \& Development in Information Retrieval}{July 8--12, 2018}{Ann Arbor, MI, USA}
	\acmBooktitle{SIGIR '18: The 41st International ACM SIGIR Conference on Research \& Development in Information Retrieval, July 8--12, 2018, Ann Arbor, MI, USA}
	\acmDOI{10.1145/3209978.3210169}
	\acmISBN{978-1-4503-5657-2/18/07}
	
	\fancyhead{}
	\maketitle

	\section{Introduction}
When people use search engines to retrieve information, acquire
knowledge or solve daily-life problems, they often are not satisfied
by a single-shot query. Instead, they will issue a series of queries
and have multiple rounds of interaction with the search engines. This is known as an information exploration process, in which each round of
interaction is a stepping stone for a user to achieve his / her final
goal.  Users' interaction with search engines is usually recorded in
search history logs. If used wisely, these search history logs can help users
preserve and recall the process of their information exploration, so
that they can re-find forgotten information or knowledge
quickly. 
A survey of experienced Web users found that people would like to use search engines to re-find online information, but often have difficulty remembering the sequence of queries they had used when they originally discovered the content in question~\cite{aula2005information}. 
In addition, studies have shown that as many as 40\% of users search queries are attempts to re-find previously encountered results~\cite{teevan2007information}.

Besides being helpful in information re-finding, search histories can also benefit collaborative search.
In collaborative search, a group of users undertake various search
subtasks, aiming to accomplish a complex collaborative task, e.g.,
planning for a trip. By seeing each other's search histories, group
members from different backgrounds can learn from each other, as they
will be looking at different aspects of the same topic / task. This
helps them form a more complete view of a certain topic or detect fake
information more effectively.

To make the best of search histories, researchers have worked on tools
that can track users' search history and visualize it in an
understandable and in-depth
presentation~\cite{morris2007searchtogether, brennan2008coordinating}.
Search logs record a searcher's explicit activities, including the
queries submitted and the answers (search results) clicked.  In
reality, such explicit activities provide only partial information
about an information exploration process.  More intellectual
activities are carried out in the searcher's mind.  Studies have shown
that a good visualization of a user search history should not only
present the explicit activities, represented by search queries and
answers, but also depict the latent information exploration process in
the searcher's mind~\cite{hearst2009search}. Such a visualization can
help users quickly re-construct the knowledge acquired in the process
of information exploration.

\begin{figure*}	
	\vspace{1.35em}
	\hspace{-3em}
	
	\subfigure[Personal View]{
		\vspace{1em}
		\hspace{-1em}
		\includegraphics[width=0.78\textwidth]{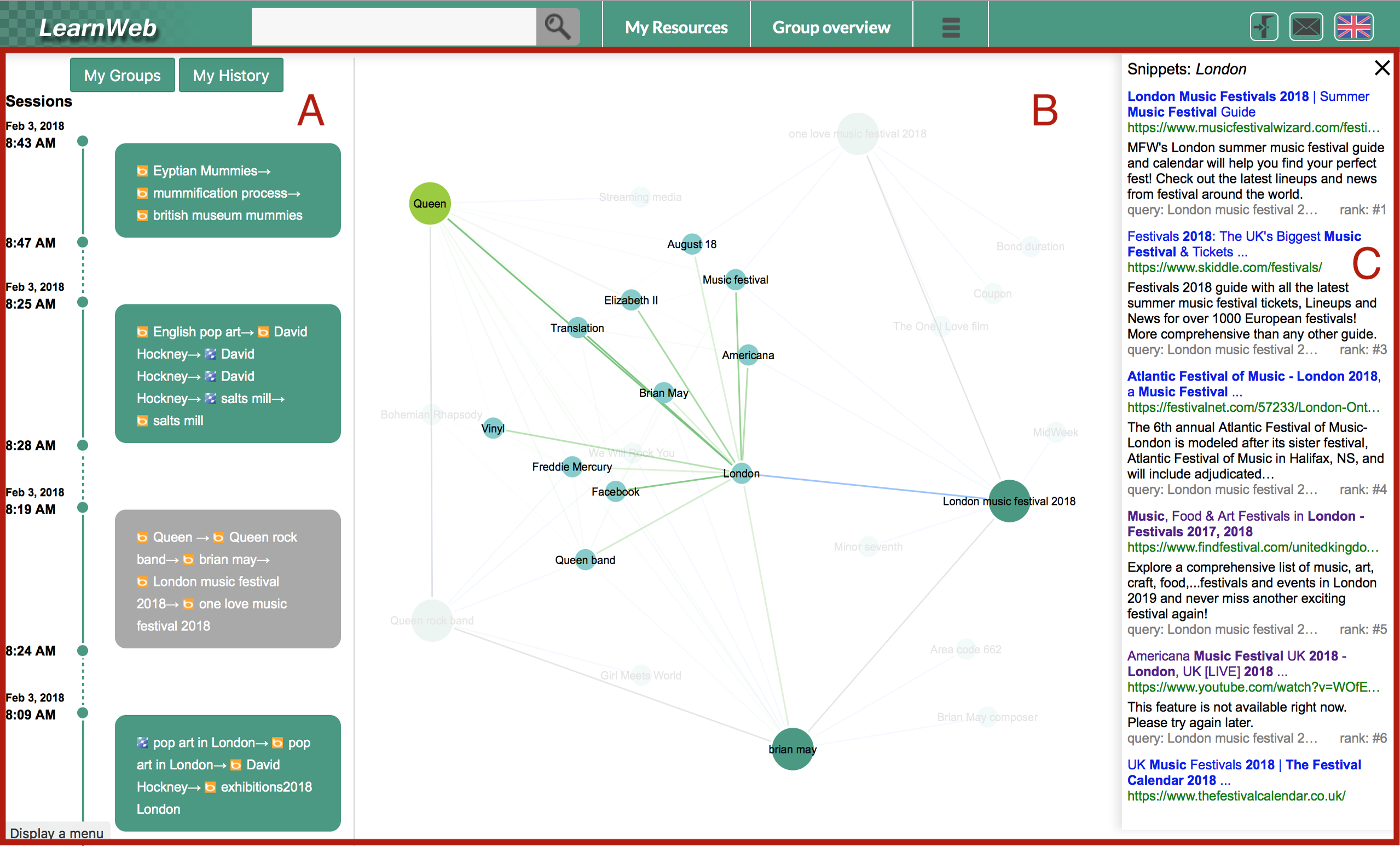}
		\hspace{-1em}
		\label{fig:c}
	}
	\vspace{1em}
	\subfigure[Group View: group session viewer of "London Attractions"]{
		\vspace{-1em}
		\includegraphics[width=0.2\textwidth]{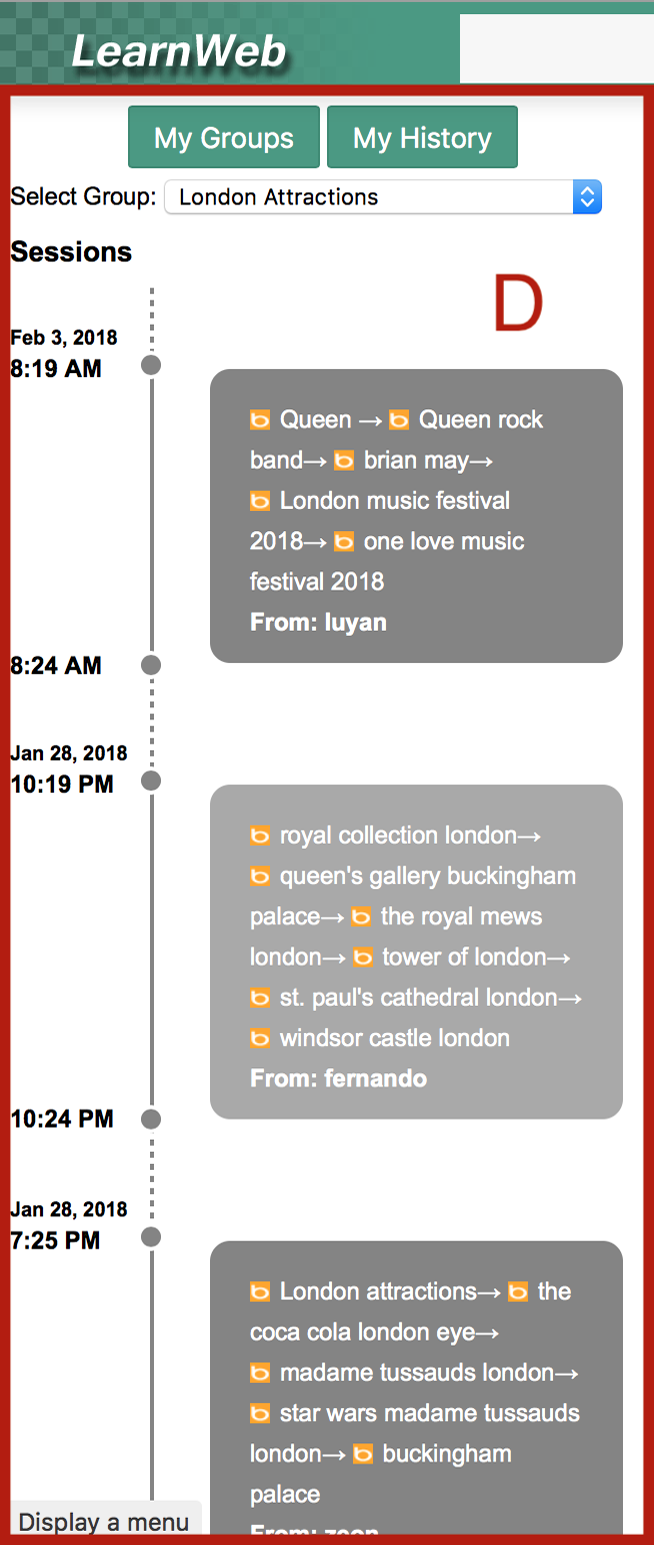}
		
		\hspace{-1em}
		\label{fig:d}
	}
	\vspace{-1em}
	\caption{\label{fig:all-the-data}Overview of LogCanvas.
		(A): personal session viewer in which queries are clustered into sessions in a most recent ordering, session "Queen $\rightarrow$...$\rightarrow$ one love music festival 2018" is selected;
		(B): knowledge-graph viewer, presenting the knowledge graph of the selected session; subgraph of node "London" is highlighted by moving the mouse over it;
		(C): snippets viewer, showing previous search results of a selected node "London";
		(D): group session viewer, in which a session from another user (fernando) is selected;
	}
	\vspace{-1em}
\end{figure*}

In this paper, we present LogCanvas, a platform for graph-based search history
visualisation.  The visualization interface is illustrated in
Figure~\ref{fig:all-the-data}. Users' search activities are clustered
into sessions (on the left). Queries of each session are embedded into
a knowledge graph to help users understand what the queries returned
as results and how they are related. This knowledge graph is
constructed based on the concepts extracted from the search result
snippets. A snippets viewer on the right helps users re-find
information of previous searches. Additional features enable users to
interact with the search histories. For instance, users can focus on a
subgraph of a certain node (e.g. "London") through a mouse hover
(Figure~\ref{fig:c}). All nodes float in the workspace, allowing users
to drag and move nodes to positions they like.

As an example, suppose an art enthusiast, who is a member of a
``London Attractions" searching group, has conducted several rounds of
searches about London. In the end, he wants to quickly review his
search history, in order to plan the trip to London.
Through the session viewer, the user can identify the sessions about
activities in London as shown in Figure~\ref{fig:c}-A.
By selecting a certain session which includes queries such as ``London
music festivals 2018'' or ``Brian May'', the user can view its
corresponding knowledge graph in the knowledge graph
viewer(Figure~\ref{fig:c}-B). All concepts in this graph are extracted
from the user's search results, enabling him to quickly grasp the
knowledge structure of his search history.
Through concepts such as "Music festival", "Americana", and "Brian
May", the user remembers that in this search session he mainly
searched about music festivals in London, specifically for
American-style music and the Queen Band.
To get details about music festivals in London in 2018, the user
clicks on "London", through the snippets viewer
(Figure~\ref{fig:c}-C), he reviews all filtered search result snippets
that are related to music festivals in London.
In order to collect more information about attractions in London, the
user can also turn to other searchers in the collaborative group
"London Attractions", to view group members' search
histories(Figure~\ref{fig:d}-D). His group members with their
different backgrounds have found interesting things he has never
thought of.  He gains insights from knowledge graphs of other group
members, acknowledging and using the suggestions from them. Meanwhile, he contributes his perspective to the group.


To realize this user interface, we applied several techniques.  First,
query logs are clustered into sessions according to time
intervals. Second, for queries of each session, related concepts and
entities are extracted from the search result snippets using the Yahoo Fast Entity Linker~\cite{blanco2015fast}. Third, correlations between the concepts
are computed using a method based on the entity co-occurrence frequency in wikipedia.

The visualisation platform can be used in different platforms, and is
currently integrated into
LearnWeb\footnote{\url{https://learnweb.l3s.uni-hannover.de/}}, an
online environment that supports collaborative sensemaking by allowing
users to share and collaboratively work on content retrieved from a
variety of web sources~\cite{tlt12, marenzi12, marenzi14}.  

In the
remainder of this paper, we briefly introduce the enabling
technologies of the user interface and how we are going to demonstrate
it in the conference.

	\section{Existing Systems}
Research on archived data visualization and information re-finding is 
relevant to LogCanvas, as it concerns preserving and visualizing
users' search histories.  

Systems such as popHistory~\cite{carrasco2017pophistory} and
Warcbase\cite{lin2014infrastructure, lin2015scaling} save users' visit
data, based on which they can extract and display the most visited
websites to users. History Viewer~\cite{segura2016history} tracks
processes of exploratory search and present users with interaction
data to enable them to revisit the steps that led to certain insights.

Information re-finding tools such as
SearchBar~\cite{morris2008searchbar} provide a hierarchical history of
recent search topics, queries, results and users' notes to help users
quickly re-find the information they have searched. Some other tools,
such as SIS (Stuff I've Seen), collect users' personal data, such as
email and docs, and offer a diary list~\cite{dumais2016stuff} to help
users quickly locate past events or visited web-pages based on
dates. Some recent work~\cite{sappelli2017evaluation,
	deng2011information} has investigated how to combine context
analysis and information re-finding frameworks to remind users about
historical events according to users' current context.

In collaborative search systems such as
Coagmento~\cite{shah2010exploring} and
SearchTogether~\cite{morris2007searchtogether}, visualization of
search history usually involves multiple users' search logs including
their search queries, bookmarks, etc. Interfaces of this kind display
search histories separately according to datatypes or categories and
support notepad functions which allow group members to share
experience.

Most of the previous visualisation tools focus primarily on the
selection of suitable data to present on the user interfaces. They
leave it to the users to re-construct short term memory and semantic
relationships. 
By contrast, the visualisation of LogCanvas not only provides a
detailed overview of search history and efficient ways to re-find
information, but also introduces a knowledge graph that helps users
connect their search activities into coherent processes of semantic
information exploration.

	\section{Our System}
\subsection{Overview}
\begin{figure}
	\vspace{-1em}
	\includegraphics[width=0.5\textwidth]{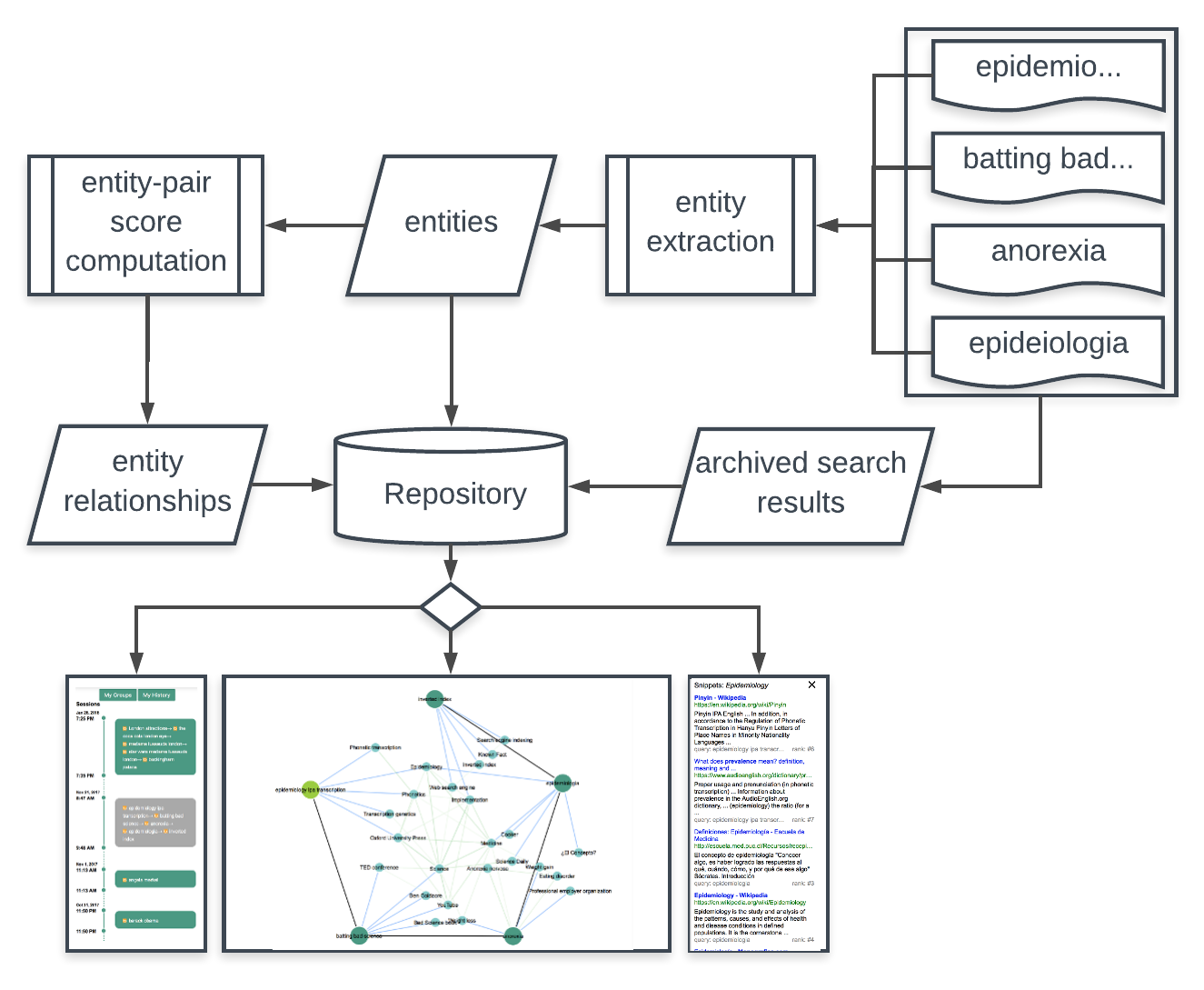}
	\label{}
	\vspace{-1em}
	\caption{\label{fig:overview}Overview of the visualization module}
	\vspace{-1em}
\end{figure}

The overview of the visualization platform is shown in
Figure~\ref{fig:overview}.  Given a users' search history log, the
following steps of data preprocessing are used to prepare the data for
final visualisation:

1) Session Segmentation -- the queries in the log are split into different sessions according to the time interval between searches(e.g. "epideiologia", "anorexia", etc.);

2) Search Result Acquisition -- the top 10 snippets of each query are
fetched from the archived search results;

3) Entity and Concept Extraction -- the most relevant entities and
concepts are extracted from the search results and form a knowledge
graph;
we use Yahoo's Fast Entity Linking
toolkit\footnote{\url{https://github.com/yahoo/FEL}}(yahooFEL)~\cite{blanco2015fast}
to extract the entities and concepts from the snippets of the top-10
search results and select the top 5 entities / concepts to add to the
knowledge graph; the selection is based on the scoring method
described in section~\ref{related_entities_scoring};

4) Edge Score Measuring -- the entity and concept nodes in the
knowledge graph are connected by edges, which represent the semantic
relationships; the edge score is computed based on the co-occurrence
of the entities and concepts in Wikipedia, as described in
section~\ref{edge_score_computation};

5) Group clustering -- a user's query session is added to the
collaborative group he/she belongs to, if any search result of the
session is tagged as useful to that group.
\subsection{Key Methodologies}
\subsubsection{Entity and Concept Extraction} \label{related_entities_scoring}
We use the Yahoo entity linking toolkit to get the candidate entities
and concepts from the top-10 search results.  To select the five most
relevant entities / concepts, we compute a quality score
$\mathit{qScore(e)}$ for each candidate $\mathit{e}$.
In general, the larger $\mathit{qScore(e)}$, the more relevant
$\mathit{e}$ to the query.  $\mathit{qScore(e)}$ is defined as
follows:
\begin{align}
\mathit{qScore_{e}} =\mathit{Freq_e}* \frac{1}{|avgFel_e|} \label{eq_1}\\
\intertext{where}
\mathit{avgFel_e}=\frac{\sum_{i=1}^{n} \mathit{fel\_score_{e_i}}}{n} \label{eq_2}
\end{align}
In eq. \ref{eq_2}, $e_i$ refers to the $i$th entity $e$ extracted from
the top-10 search result snippets. $\mathit{fel\_score_{e_i}}$ is a
negative value returned by yahooFEL.  It represents the confidence of
yahooFEL in that $e_i$ is a relevant entity to the query.
$\mathit{n}$ is the count of occurrence of an entity $\mathit{e}$ in
the top-10 search result snippets.
We sum the $\mathit{fel\_score_{e_i}}$ from different snippets and get
an average $\mathit{avgFel_e}$, which represents the confidence of
yahooFEL in that $e$ is relevant to the entire session.
In eq. \ref{eq_1}, $\mathit{Freq_{e}}$ is the frequency of
$\mathit{e}$ occurring in the top-10 result snippets.  We include this
to favor more frequently occurring entities in the top-10 results.
In addition, we remove candidate entities that have a word length less
than 4, as they are likely to be stop words.

\subsubsection{Edge Score Computation} \label{edge_score_computation}
An edge in the knowledge graph represents the semantic association
between two entities / concepts.  For entities in one session, we
enumerate all possible entity pairs and count the co-occurrence of
each pair in Wikipedia.  The co-occurrence count is the number of
documents returned from the Wikipedia index built from
Solr\footnote{\url{http://lucene.apache.org/solr/}}, using a boolean
query with the two entities as phrase terms (e.g., ``computer science"
AND ``information retrieval").  We assume that the more frequently two
entities co-occur in Wikipedia articles, the more correlated they are.

Some strongly correlated entities can still get a low co-occurrence
count when one of the entities is not common in wikipedia. On the
other hand, some entity pairs with high co-occurrence count (i.e.,
``human" and ``animal") are common sense correlations uninteresting to
users.  Moreover, correlation of the same entity pair can have
different meanings in different search sessions.  For example, the
relationship between ``apple" and ``toolkit" in the session of
``computer" is different from that of ``fruit cultivation".

Therefore, to make the entity relationships meaningful targeting a
certain session, we normalize the co-occurrence counts using a
non-linear function.  For each entity pair $\mathit{(e_{i}, e_{j})}$
of a session, the $\mathit{eScore_{(e_{i}, e_{j})}}$is defined as
follows:
\[
\mathit{eScore_{(e_{i}, e_{j})}} =
\begin{cases}
1 - \frac{ \lambda}{ \lambda + \max_{} \{  \mathit{C_{(e_{i}, e_{j})}}  \}}
& \quad \text{if } \max_{} \{  \mathit{C_{(e_{i}, e_{j})}}  \} > 1000 \\
\frac{ \mathit{C_{(e_{i}, e_{j})}}}{   \max_{} \{  \mathit{C_{(e_{i}, e_{j})}}  \} }
& \quad \text{otherwise }
\end{cases}
\]
where $\mathit{C_{(e_{i}, e_{j})}}$ means how many times two entities
co-occur in wikipedia and $\max_{} \{ \mathit{C_{(e_{i}, e_{j})}} \}$
is the entity pair with the largest co-occurrence count in a session.
With the non-linear function, even when there is an entity pair with
an extremely large co-occurrence count(>1000), the edge scores of the
other entity pairs can still be significant. This makes sure that all
relevant edges are visible in the graph visualisation. We empirically
set $\lambda$ to 50 in the normalization.

\subsubsection{Data Collection}
To collect users' search histories, whenever a user submits a query to
the platform (i.e., LearnWeb), we record the query, the search
objective (text, image, video) and the search service provider (bing,
flickr, youtube, etc.) in the history log, and annotate them with a
timestamp. All top search results viewed by the user are also stored
based on the click or save information (when a result is saved to a
group in LearnWeb).
These preprocessing steps are then performed offline once a certain
amount of log has been accumulated - we run the edge score computation
script at the end of each day.
All results are stored in a relational database format in a MariaDB
repository, so they can be quickly retrieved during visualization.

	\section{Demonstration}

In the demonstration, we will mainly show how LogCanvas can help user re-find information and how it can benefit collaborative search. 

On our LearnWeb platform, we have collected a large number of user search histories. The histories include those of individuals who used LearnWeb to search and explore learning resources on the Web. They also include histories of collaborative search processes, in which a group of users studied a topic together and shared their findings. 
In the demonstration, we will visualize these search histories in LogCanvas. We will let the audience interact with the search histories to show how our visualization can help them quickly understand the search processes. 

We will also provide a number of test scenarios for the audience to try out our system. They will first perform some search tasks using LearnWeb. During this process, audience can create or join a collaboratively searching group.  
We will generate the visualization of their search histories on the fly, and demonstrate how accurately LogCanvas can visualize their information exploration processes.

At this moment, an online demonstration of
LogCanvas\footnote{\url{http://learnweb.l3s.uni-hannover.de/lw/searchHistory/entityRelationship.jsf}}
is
accessible using the demo account (username: \textit{luyan}, password:
\textit{test}).

	\begin{acks}
	This research has been supported in part by project ALEXANDRIA which is funded by  European Research Council under the EU 7th Framework Programme (FP7/2007-2013) / ERC 339233.
	\end{acks}
	
	\begin{flushleft}
		
	\end{flushleft}
	
	\bibliographystyle{acm-ref}
	\bibliography{my}
	
\end{document}